\documentclass[letter,twocolumn]{jpsj2} %% for letters
\usepackage{graphicx}
\usepackage{color}
\title
{Magnetic properties of the extended periodic Anderson model}

\author
{
Akihisa \textsc{Koga},$^{1,2}$ Norio \textsc{Kawakami},$^{1}$ 
Robert \textsc{Peters}$^2$ and Thomas \textsc{Pruschke}$^2$
}

\inst{
$^1$Department of Physics, Kyoto University, Kyoto 606-8502, 
Japan\\
$^2$Institut f\"ur Theoretische Physik Universit\"at G\"ottingen,
G\"ottingen D-37077, Germany
}

\abst{
We study magnetic properties of the extended periodic Anderson model, 
which includes electron correlations within and between itinerant 
and localized bands.
By combining dynamical mean-field theory with the numerical 
renormalization group
we calculate the sublattice magnetization and the staggered
 susceptibility to determine the phase diagram 
in the particle-hole symmetric case.
We find that two kinds of magnetically ordered states compete with the
Kondo insulating state at zero temperature, which induces 
non-monotonic behavior in the temperature-dependent magnetization.
It is furthermore clarified that a novel magnetic metallic state is 
stabilized at half filling by 
the competition between Hund's coupling and the hybridization.
}

\kword{periodic Anderson model,
dynamical mean field theory, numerical renormalization group}

\begin{document}
\maketitle

%%%%%%%%%%%%%%%%%%%%%%%%%%%%%%%%%%%%%%%%%%
%\section{Introduction}
%%%%%%%%%%%%%%%%%%%%%%%%%%%%%%%%%%%%%%%%%%

Strongly correlated electron systems with degenerate orbitals 
have attracted great interest. 
One of the popular examples is the manganite $\rm (La, Sr)MnO_3$.\cite{Urushibara}
In this system, itinerant electrons in the $e_{g}$ band 
are coupled to localized electrons in the $t_{2g}$ band
through Hund's coupling, leading to a competition between
antiferromagnetic (AF) correlations and
%compete with 
the double exchange ferromagnetic correlations.
This yields a complex phase diagram with
various types of ordered ground states. 
Other interesting examples are $\rm (Ca, Sr)_2RuO_4$\cite{Nakatsuji} and 
$\rm La_{n+1}Ni_nO_{3n+1}$.\cite{Sreedhar,Kobayashi} 
In these compounds, the chemical substitution or the change in
temperature is suggested to trigger  
an orbital-selective Mott transition,\cite{Anisimov,KogaOSMT}
where some of the orbitals become localized by electron correlations,
while the others still remain itinerant.
It is also proposed that in these compounds,
localized and itinerant electrons are hybridized with each other, 
inducing heavy fermion or bad metal behavior at low temperatures.\cite{KogaV,Medici,previous,Buneman}
%Such a scenario has also been suggested in the heavy fermion compound
%$\rm LiV_2O_4$.

An important point in the above compounds is that the localized and
itinerant bands in the $d$-orbital and their correlations
play a crucial role in stabilizing the magnetically ordered state or heavy
fermion state.
Generally speaking, in the systems with localized bands,
 the hybridization together with local electron
correlations screen spins, leading to 
heavy fermion behavior with the large
density of states (DOS) around the Fermi level, the so-called Kondo effect. \cite{grewe_ssc,Grewe91,Hewson}
In contrast, Hund's coupling
% in each orbital makes 
enforces parallel spins in different orbitals,
%parallel, 
enhancing 
%where 
magnetic correlations, as discussed for the manganites.\cite{Zener,Anderson,Kubo,Furukawa}
Therefore, an interesting question arises how robust 
the nonmagnetic ground state is in systems with localized and itinerant
bands.
In a previous paper,\cite{previous} 
we have investigated the extended periodic Anderson model (EPAM) to
clarify how 
the Kondo and Mott insulating states compete
with the metallic state in the paramagnetic case.
However, a magnetic instability has not been discussed so far,
which may be important to understand low-temperature properties 
in real materials such as some transition metal oxides and $f$-electron
systems.
%Therefore, it is necessary to study how antiferromagnetically 
%ordered states with some configurations compete with the Kondo
%insulating state.
Furthermore, the competing interactions may lead to nontrivial
behavior in the magnetically ordered state.
Therefore, it is highly desirable to clarify 
the magnetic properties in the system with localized and
itinerant electrons.

%%%%%%%%%%%%%%%%%%%%%%%%%%%%%%%%%%%%%%%%%%%%%%%%%%%
%\section{Model and Method}
%%%%%%%%%%%%%%%%%%%%%%%%%%%%%%%%%%%%%%%%%%%%%%%%%%%
For this purpose, 
we consider a correlated electron system
which is described by the following Hamiltonian as,
\begin{eqnarray}
{\cal H}&=&H_t + \sum_i H_{loc}^{(i)},\label{eq:model}\\
H_t&=&\sum_{\langle ij\rangle \alpha\sigma}
\left[ t_{ij}^{(\alpha)}-\mu \delta_{ij}\right]
c_{i\alpha\sigma}^\dag c_{j\alpha\sigma},\nonumber\\
H_{loc}^{(i)}&=&V\sum_{\sigma}\left(
c_{i1\sigma}^\dag c_{i2\sigma}+c_{i2\sigma}^\dag c_{i1\sigma}
\right)\nonumber\\
&+&\sum_{\alpha} U_\alpha n_{i\alpha\uparrow} n_{i\alpha\downarrow}
+
\sum_{\sigma\sigma'}\left(U'-J\delta_{\sigma\sigma'}\right)
n_{i1\sigma}n_{i2\sigma'}\nonumber\\
&-&J
\sum_\sigma c_{i1\sigma}^\dag c_{i1\bar{\sigma}}
c_{i2\bar{\sigma}}^\dag c_{i2\sigma}
-J
\sum_\alpha
c_{i\alpha\uparrow}^\dag c_{i\alpha\downarrow}^\dag
c_{i\bar{\alpha}\uparrow} c_{i\bar{\alpha}\downarrow},\nonumber
\end{eqnarray}
%%%%%%%%%%%%%%%%%%%%%%%%
where $c_{i\alpha\sigma}^\dag (c_{i\alpha\sigma})$
creates (annihilates) an electron
with spin $\sigma(=\uparrow, \downarrow)$ and band
index $\alpha(=1, 2)$ at the $i$th site,
and $n_{i\alpha\sigma}=c_{i\alpha\sigma}^\dag c_{i\alpha\sigma}$.
For the band $\alpha$,
$t_{ij}^{(\alpha)}$ represents the transfer integral,
$V$ the hybridization between bands.
The intra-band and inter-band Coulomb interactions are described
by $U_\alpha$ and $U'$, while $J$ denotes Hund's coupling. Finally, 
$\mu$ is the chemical potential.
%The structure of the model Hamiltonian is schematically shown in Fig. 
%\ref{fig:model}.

To investigate the correlated electron
system with one band localized and the other itinerant,
we set the hopping integral for the $\alpha=2$ band to
$t_{ij}^{(2)}=0$, for simplicity.
Then this model is regarded as the EPAM
with not only intraband interactions but also interband ones.
Here, to discuss magnetic properties,
we make use of dynamical mean-field theory (DMFT).\cite{Metzner,Muller,Georges,Pruschke}
%, which 
%has been developed in several groups.
%and has successfully been applied to correlated electron systems.
% such as
%the single-band Hubbard model,
%\cite{Caffarel,OSakai,single1,single2,single3,single4,BullaNRG}
%the two-band Hubbard model
%\cite{2band1,2band2,Koga,Momoi,OnoED,multi,Ono4,Pruschke2,KogaOSMT,OSMT}
%or the periodic Anderson model.
%\cite{PAM,Mutou,Saso,pam_pr,Sato,Ohashi,Medici}
In DMFT, a lattice model is mapped to an effective quantum impurity,
where local electronic correlations are taken into account exactly.
The requirement that the site-diagonal lattice Green function 
is equal
to that of the effective quantum impurity then leads to a self-consistency
condition for the parameters entering the impurity problem.
This treatment is formally exact in infinite spatial dimensions and
even for three dimensions reliable results are obtained if 
non-local correlations are allowed to be ignored.

When an AF instability is treated 
in the framework of DMFT,\cite{Chitra,Zitzler,Momoi,PetersNew}
%we divide the lattice into two sublattices.\cite{Chitra}
%Since the self-energy between sublattices does not appear,
%dynamical properties in the system can be described in terms of 
%the effective impurity model for each sublattice.
the self-consistency equation for the sublattice $\gamma [=(A, B)]$
is represented as,
\begin{eqnarray}
\left[\hat{\cal G}_{0\; \gamma \; \sigma}^{-1}(z)\right]_{11}=
z+\mu-\left(\frac{D}{2}\right)^2
\left[\hat{G}_{loc \; \bar{\gamma} \; \sigma}(z)\right]_{11},
\label{eq:self}
\end{eqnarray}
where 
$\hat{\cal G}_0$ is the non-interacting Green function for the effective
impurity model and $\hat{G}_{loc}$ the local Green function. 
Here, we have used the Bethe lattice 
with infinite coordination, where 
$D$ is the half bandwidth for the bare itinerant band.
Note that the self-consistency equation is represented only by
one component of the Green function.\cite{Sato,previous,Schork}
Therefore, we can introduce the effective impurity model, where
one of the impurity bands connects to the effective bath.
The corresponding hybridization function is then defined by 
%\begin{eqnarray}
$\Delta_{\gamma\sigma} = \left(\frac{D}{2}\right)^2 
\left[\hat{G}_{loc \; \bar{\gamma} \;\sigma} (z)\right]_{11}.$
%\end{eqnarray}
To solve the effective impurity model quantitatively, we use the numerical
renormalization group (NRG)\cite{NRG,nrg_rmp} as an impurity solver.\cite{OSakai,BullaNRG}
This allows us to access low energy properties,
which are particularly important in the Kondo insulating state.
Details of the method can be found in literature.\cite{anders,Peters,weichselbaum:06}
The hybridization function $\Delta_{\gamma\sigma}$ 
should be determined self-consistently through
the DMFT condition eq. (\ref{eq:self}).

\begin{figure}[htb]
\vspace{-.6cm}
\begin{center}
\includegraphics[width=7.6cm,clip]{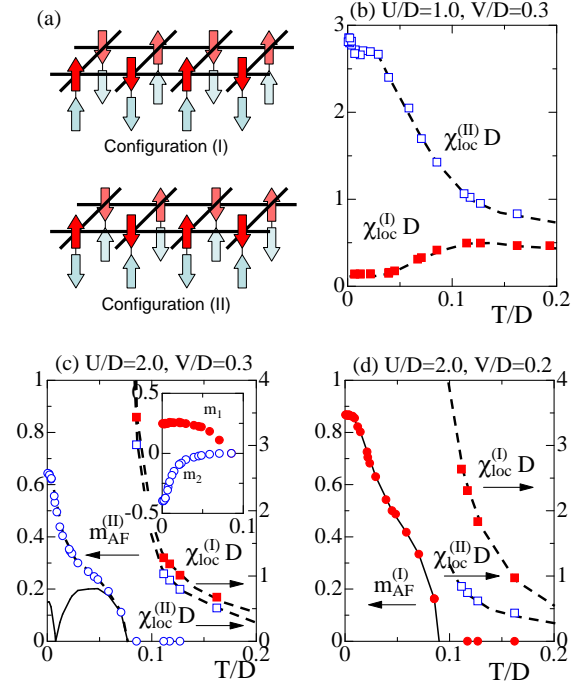}
\end{center}
\vskip -4mm
\caption{(a) Possible spin configurations.
(b), (c) and (d) show the staggered susceptibilities and the
 magnetizations as a function of the temperature $T$.
Solid lines represent $|m_{AF}^{(I)}|$ and dashed lines are guide to eyes.
Solid and open symbols represent the results
for the configuration (I) and (II). 
}
\vskip -4mm
\label{fig:fig1}
\end{figure}

In the half-filled system without frustration, 
the AF ground state is stabilized if
intersite correlations are large enough.
At each site, the ferromagnetic Hund's coupling competes with 
the effective AF exchange coupling induced by 
the Coulomb interaction together with the hybridization.\cite{previous}
Therefore,
two possible spin configurations are naively expected for the AF states,
which are schematically shown in Fig. \ref{fig:fig1} (a).
A magnetization for each configuration may be given as
$m_{AF}^{(I, II)} = m_1 \pm m_2$
where $m_\alpha=\sum_i(-1)^{P_i}
(n_{i\alpha\uparrow}-n_{i\alpha\downarrow}) /(2N)$, 
$P_i= 0 (1)$ for $i \in A (B)$, and $N$ is the total number of sites.
Note that these magnetizations are not ordinary order parameters
characterizing the AF (I) and (II) states 
since they should be finite in both states.
Nevertheless, we can distinguish between these states:
When $|m_{AF}^{(I)}|-|m_{AF}^{(II)}|>0 (<0)$,
the AF (I) [(II)] is realized, and
a first-order phase transition (crossover)
between the two states occurs at $T=0 (T\neq 0)$, where $T$ is the temperature.
Furthermore, $m_{AF}^{(I)}$ is an important quantity since
it characterizes the magnetization of ions in real materials and 
can be observed in inelastic neutron scattering experiments.
Here, to discuss how magnetic fluctuations develop at low temperatures,
we calculate the magnetization and the staggered susceptibility 
$\chi_{loc} (=\partial m_{AF}/\partial 
h_{AF})$, where $h_{AF}$ is the staggered magnetic field.
The susceptibility is obtained as the slope
%derivative
of  the sublattice magnetization 
for a tiny field $h_{AF}/D \sim 0.002$,
which gives a good estimate
except in the vicinity of the critical temperature.

In this paper, we focus on the half-filled EPAM
to discuss magnetic properties.
In particular, we fix the ratio $\lambda=J/U=0.1$ and $U=U'+2J$ to clarify 
how the competition between Hund's coupling and the
hybridization affects the magnetic phase diagram at low temperatures.
The effects of hole doping and/or the magnetic field, 
which may yield a complex phase diagram
including several types of magnetically ordered states,\cite{Ohashi,Milat} 
are also interesting problems.
However, these are out of the scope in this paper, 
and will be discussed elsewhere.

%%%%%%%%%%%%%%%%%%%%%%%%%%%%%%%%%%%%%%%%%%%%%%%%%%%
%\section{Magnetic properties at half filling}
%%%%%%%%%%%%%%%%%%%%%%%%%%%%%%%%%%%%%%%%%%%%%%%%%%%
In Figs. \ref{fig:fig1} (b)-(d), 
we show the results obtained by the DMFT with the NRG.
%%%%%%%%%%%%%%%%%%%%%%%%%%%%%%%%%%%%%%%%%%%%%%%%%%%%%%%%%%%%%%%%%%
%%%%%%%%%%%%%%%%%%%%%%%%%%%%%%%%%%%%%%%%%%%%%%%%%%%%%%%%%%%%%%%
When $U/D=1.0$ and $V/D=0.3$, no singularity appears in both staggered
susceptibilities, as shown in Fig. \ref{fig:fig1} (b).
This suggests that  
the non-magnetic Kondo singlet ground state
is realized at low temperatures. 
We also find that 
magnetic fluctuations for the configuration (II) are enhanced
at low temperatures and the system is close to the AF (II) state.
In fact, when the parameters are slightly changed, 
the AF (II) state appears at low temperatures.
The staggered susceptibility diverges 
at a critical temperature $T_N$
and a spontaneous magnetization $m_{AF}^{(II)}$
appears below $T_N$,
as shown in Fig. \ref{fig:fig1} (c).
%This clarifies that the AF (II) phase is stabilized 
%below $T_N$ when $U/D=2.0$ and $V/D=0.3$.
It is also found that a shoulder structure appears in
the temperature-dependent magnetization.
This implies that magnetic correlations for each band 
are not enhanced at the same temperature (see also the inset).
This interesting feature will be discussed later.
%Furthermore, one finds that the susceptibilities for both configurations
%are almost equally enhanced around the critical temperature.
%These behaviors will be discussed later.
On the other hand, when the Coulomb interaction and Hund's
coupling are relatively large ($U/D=2.0$ and $V/D=0.2$), 
$\chi_{AF}^{(I)}$ diverges 
at the critical temperature and
$m_{AF}^{(I)}$ appears at lower temperatures, 
as shown in Fig. \ref{fig:fig1} (d).
The AF (I) state is then realized in the ground state.

\begin{figure}[htb]
\vspace{-.8cm}
\begin{center}
\includegraphics[width=6.5cm]{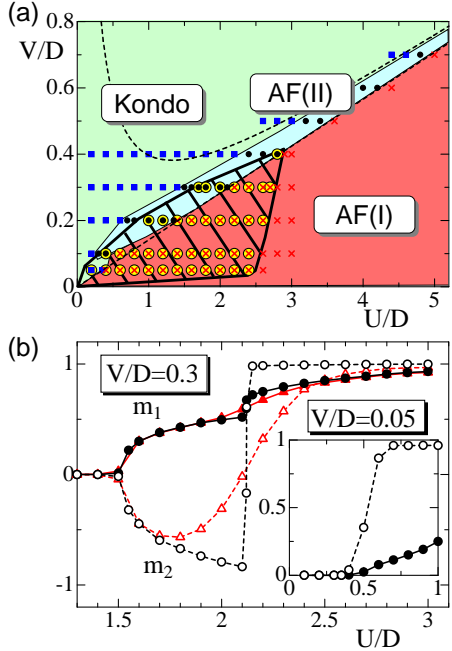}
\end{center}
\vskip -4mm
\caption{
(a) The phase diagram of the EPAM at $T/D=4.6\times 10^{-4}$.
Solid squares, crosses and solid circles represent
 the Kondo singlet, AF (I) and AF (II) phases, respectively.
%Open circles indicate the metallic state, which is realized in the
% shaded area.
Open circles indicate the metallic state realized 
in the shaded area,
which will be discussed (Fig. \ref{fig:fig4}).
The phase boundaries are guide to eyes. Dashed line is the phase
 boundary obtained from the strong coupling limit.
(b) The magnetization for each band when $V/D=0.3$. 
Circles and triangles represent the results at 
$T/D=4.6\times 10^{-4}$ and $0.015$, respectively.
The inset shows the results for $V/D=0.05$. 
}
\vskip -4mm
\label{fig:fig2}
\end{figure}

By performing similar calculations for various model parameters, 
we end up with the low-temperature phase diagram at half filling, 
shown in Fig. \ref{fig:fig2} (a).
%When $V/D \gg U/D$, 
%the hybridization together with local electron correlations stabilizes
%the Kondo insulator,
%as shown in Fig. \ref{fig:mag} (d).
%On the other hand, in the large $U$ case,
%strong electron correlations induce the magnetically ordered states. 
%In fact, the large Hund coupling 
%stabilizes the AF (I) phase in a wide region, 
%and the AF (II) phase appears 
%between the Kondo and the AF (I) phases.
To discuss the phase transitions between these states in detail,
we show in Fig. \ref{fig:fig2} (b)
the staggered magnetization for each band when $V/D=0.3$.
%%%%%%%%%%%%%%%%%%%%%%%%%%%%%%%%%%%%%%%%%%%%%%%%%%%%%%%%%%%%%%%%%%
%%%%%%%%%%%%%%%%%%%%%%%%%%%%%%%%%%%%%%%%%%%%%%%%%%%%%%%%%%%%%%%
%First, we discuss the low-temperature magnetic properties.
In the case of small $U/D$, the interband hybridization screens the local
moments, and
no magnetizations appear in both bands.  
Therefore, in the region $(U/D) < (U/D)_{c1}[\sim 1.5]$ 
the paramagnetic Kondo phase is realized, as shown in
Fig. \ref{fig:fig2} (b).
The increase in the interaction induces
magnetic moments for both bands with opposite signs.
This implies that a continuous phase transition occurs 
to the AF (II) phase.
Further increase in the interaction 
leads to a jump singularity in the temperature-dependent magnetization, 
at which the sublattice magnetizations become parallel.
The first order phase transition then drives the system 
to the AF (I) state at $(U/D)_{c2}\sim 2.1$.
It is also found that the jump singularity 
vanishes when the temperature is slightly increased,
as shown in Fig. \ref{fig:fig2} (b).
Therefore, the phase transition between two AF states is present at zero
temperature only, while a crossover occurs at finite temperatures.

\begin{figure}[htb]
\vspace{-8mm}
\begin{center}
\includegraphics[width=6cm]{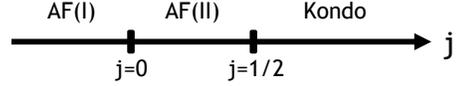}
\end{center}
\vskip -5mm
\caption{Ground-state phase diagram for the Kondo necklace model. 
}
\vskip -6mm
\label{fig:spin}
\end{figure}

The competition between these phases
may be explained by considering the strong coupling limit 
$(U\rightarrow \infty)$.
The system is then reduced to the Kondo necklace Heisenberg model as,
$H=J_{inter}\sum_{ij}{\bf S}_{i1}\cdot {\bf S}_{j1}+ 
J_{intra}\sum_i {\bf S}_{i1}\cdot {\bf S}_{i2}$, 
where ${\bf S}_{i\alpha}=\sum_{ss'}\frac{1}{2}c_{i\alpha s}^\dag 
\sigma_{ss'} c_{i\alpha s'}$,
$J_{inter}$ is the intersite effective exchange coupling,
and the intrasite one is represented by $J_{intra}$, 
instead of $J$.
We note that $J_{inter}(=4t^2/U)$ is always positive, while
$J_{intra}$ depends on the interactions and
the hybridization. 
Its magnitude is given by the lowest singlet-triplet gap
of the local Hamiltonian $H_{loc}$ as $J_{intra}=\Delta E 
[= U(\sqrt{\lambda^2+4 (V/U)^2}-3\lambda)]$.
In the model, three distinct phases appear in the phase diagram, 
which is schematically shown in Fig. \ref{fig:spin}.
%%%%%%%%%%%%%%%%%%%%%%%%%%%%%%%%%%%%%%%%%%%%%%%%%%%%%%%%%%%%%%%%%%
%%%%%%%%%%%%%%%%%%%%%%%%%%%%%%%%%%%%%%%%%%%%%%%%%%%%%%%%%%%%%%%
When $j(= zJ_{intra}/J_{inter}) \gg 1$, the Kondo singlet phase is
stabilized, where
$z$ is the coordination number.
An increase in the intersite coupling $J_{inter}$ 
enhances AF correlations and
a second-order phase transition, at last,
occurs to the AF (II) phase at 
a critical value $j_c(=1/2)$.
The phase boundary obtained from the strong coupling limit,
which is given by $(V/D)_c=(U/4D)\sqrt{(4\lambda+(D/U)^2)(8\lambda+(D/U)^2}$,
is consistent with that in the EPAM, as shown in Fig. \ref{fig:fig2} (a).
On the other hand, the AF (I) and (II) phases are separated by the
condition $j=0 [V/D = \sqrt{2} \lambda (U/D)]$,
where the localized spins ${\bf S}_{i2}$ are completely decoupled and 
the phase transition never occurs.
In contrast, two bands are coupled through the hybridization 
$V$ in the EPAM, 
leading to a first-order phase transition.
The corresponding phase boundary is in good agreement 
with the condition $j=0$.
When $V/D=0.05$ and $T/D=4.6\times 10^{-4}$,
we could not find the AF (II) phase 
between AF (I) and Kondo insulating phases,
as shown in the inset of Fig. \ref{fig:fig2} (b).
This is consistent with the fact that in the weak coupling region
the energy scale of magnetic correlations is fairly small and 
the AF state is stable only at very low temperatures.
Therefore, we believe that in the ground-state magnetic phase
diagram,  
the AF (II) phase always appears
between the AF (I) and Kondo phases.

Next, we discuss the finite-temperature magnetic properties
such as the shoulder structure in the magnetization shown before. 
When one concentrates on the local Hamiltonian $H_{loc}$, 
the low-lying singlet and triplet states can be 
considered to be four-fold degenerate down to a certain temperature 
$T^*=|\Delta E|$.
When $T^*$ is sufficiently low, 
an intersite exchange stabilizes 
the magnetically ordered state, where the itinerant band is 
almost fully polarized
% magnetized 
while {\it nearly free localized spins} appear in the other.
This reveals that
orbital-selective like features appear in the intermediate 
temperature $T^*<T<T_N$ and
magnetic correlations in the localized band are enhanced
below $T^*$. 
Such behavior is clearly found in the
vicinity of the phase boundary between AF (I) and (II) states,
where $T^* \ll T_N$. 
In fact, when $U/D=2.0$ and $V/D=0.3$, 
$(T^*, T_N) \sim (0.032D, 0.075D)$
and the shoulder structure appears in the temperature-dependent 
magnetization, as shown in 
the inset of Fig. \ref{fig:fig1} (c).
Furthermore, we find
nontrivial behavior in the observable quantity $|m_{AF}^{(I)}|$.
Namely, when decreasing the temperature, 
it once vanishes at a certain temperature 
below $T^*$, and is induced again, shown as the solid line
in Fig. \ref{fig:fig1} (c).
This non-monotonic behavior 
may be observed in the intensity of the magnetic peak 
in the inelastic neutron scattering experiments
for some transition-metal oxides with localized and itinerant bands
such as $\rm Ca_{2-x} Sr_x RuO_4 (0<x<0.5)$,
which then allows us to discuss the competition between Hund's coupling and
the hybridization at each site.

\begin{figure}[htb]
\vspace{-.6cm}
\begin{center}
\includegraphics[width=7.cm]{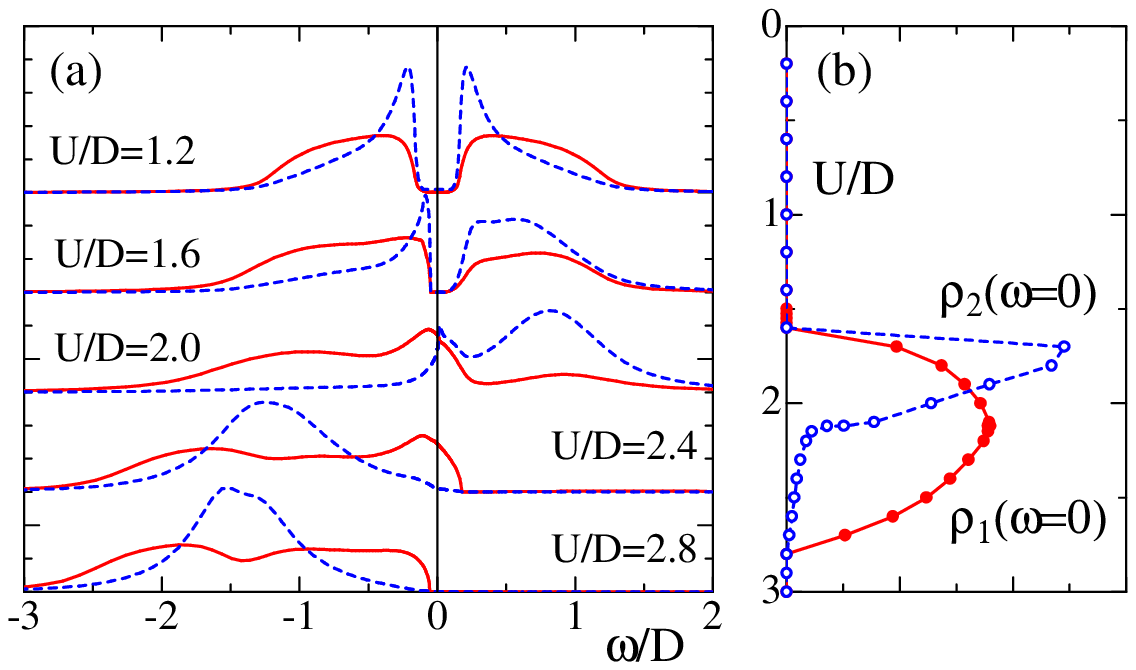}
\end{center}
\vskip -4mm
\caption{
(a) DOS in the system with the majority spin in the itinerant band 
when $V/D=0.3$ and $U/D=1.2, 1.6, 2.0, 2.4, 2.8$ at $T/D=4.6\times 10^{-4}$.
(b) DOS at the Fermi level as a function of $U/D$.
Solid (dashed) lines represent the results for the itinerant (localized)
band.}
\vskip -4mm
\label{fig:fig4}
\end{figure}

We would like to mention another remarkable feature for 
the metal-insulator (MI) transition at half filling.
In infinite dimensions, the Hubbard and  the Kondo lattice model 
have a chance
to realize the MI transition 
without magnetism when the system is strongly frustrated.
However, in the system without frustration, the magnetically ordered
state is more stable than the paramagnetic metallic state 
and the Mott insulating state at zero temperature, 
where the charge gap always opens around the Fermi level.
\cite{Zitzler,PetersNew}
The EPAM treated here is not frustrated, but has 
 competing interactions between the orbitals at each site,
which may be referred to as {\it orbital frustration}.
Therefore, it is not trivial whether the ground state is always
insulating under orbital frustration.
To clarify this, we show the DOS in Fig. \ref{fig:fig4} (a).
%%%%%%%%%%%%%%%%%%%%%%%%%%%%%%%%%%%%%%%%%%%%%%%%%%%%%%%%%%%%%%%%%%
%%%%%%%%%%%%%%%%%%%%%%%%%%%%%%%%%%%%%%%%%%%%%%%%%%%%%%%%%%%%%%%
When $U/D \leq 1.5$, the Kondo insulating state is realized, where
the hybridization gap clearly appears in the localized band.
In the case $U/D \geq 2.8$, the structures in the DOS in both bands
appear far from the Fermi level, where the polarized ground state is 
realized with the configuration (I).
On the other hand, when the system approaches the intermediate region,
spectral weight is built up around the Fermi level.
In fact, Fig. \ref{fig:fig4} (b) clearly shows that
a finite DOS at each band appears in the intermediate region, 
although the $\alpha=2$ band is almost localized in the case $(2.4<U<2.8)$.
This fact implies that a metallic phase appears between
the different insulating states. 
Since the DOS is continuously changed at 
the metal-insulator transition points,
a drastic change was not found in the magnetization
shown in Fig. \ref{fig:fig2} (b).
In our phase diagram,
this magnetic metal is stable in the weak coupling region
shown as the shaded area in Fig. \ref{fig:fig2} (a).
We recall that when $V=0 (U=0)$, the system is reduced to
the Kondo lattice (conventional periodic Anderson) model, where
the AF (Kondo) insulator is always realized except for the decoupled
limit $U=V=0$.
Therefore, we claim that 
the competition between Hund's coupling and the
hybridization introducing a kind of orbital frustration plays an important role 
in stabilizing the magnetic metal.
This is in contrast to the other correlated
systems such as the Hubbard model and the Kondo lattice model,
where the magnetic insulating state is always realized, as mentioned above.

%%%%%%%%%%%%%%%%%%%%%%%%%%%%%%%%%%%%%%%%%%%%%%%%
%\section{Summary}
%%%%%%%%%%%%%%%%%%%%%%%%%%%%%%%%%%%%%%%%%%%%%%%%
In summery, 
we have investigated the extended periodic Anderson model by combining
the dynamical mean-field theory with the numerical renormalization
group. We have discussed the magnetic properties at low temperatures to
clarify how the magnetically ordered states compete with the Kondo
insulating state. Furthermore, we have for the first time found the
magnetic metallic state in the half-filled system
without lattice frustration, which is instead stabilized by {\it orbital
frustration}.

%%%%%%%%%%%%%%%%%%%%%%%%%%%%%%%%%%%%%%%%%%%%%%%%%%%%%%%%
%\section*{Acknowledgment}
%%%%%%%%%%%%%%%%%%%%%%%%%%%%%%%%%%%%%%%%%%%%%%%%%%
This work was partly supported by a Grant-in-Aid from the Ministry 
of Education, Science, Sports and Culture of Japan 
[19014013 (N.K.), 17740226 (A.K.)]
and the German Science Foundation (DFG) through the
collaborative research grant SFB 602 (T.P., A.K.) and 
the project PR 298/10-1 (R.P.)  
A.K. would
in particular like to thank the SFB 602 for its support and hospitality during
his stays at the University of G\"ottingen.
Part of the computations were done at the Supercomputer Center at the 
Institute for Solid State Physics, University of Tokyo
and Yukawa Institute Computer Facility.

%%%%%%%%%%%%%%%%%%%%%%%%%%%%%%%%%%%%%%%%%%%%%%%


\begin{thebibliography}{99}
%% The number "99" means that this list has more than nine items.

\bibitem{Urushibara}
A. Urushibara, {\it et al.},
%A. Urushibara, Y. Moritomo, T. Arima, A. Asamitsu, G. Kido, and Y. Tokura,
Phys. Rev. B {\bf 51}, 14103 (1995);
Y. Moritomo, {\it et al.},
%Y. Moritomo, A. Asamitsu, H. Kuwahara, and Y. Tokura, 
Nature {\bf 380}, 141 (1996).

%% LiV2O4 %5
%\bibitem{Kondo}
%S. Kondo, {\it et al.}, Phys. Rev. Lett. {\bf 78}, 3729 (1997).

% LiV2O4 theory %
%\bibitem{Kaps}
%H. Kaps, N. B\"uttgen, W. Trinkl, A. Loidl, M. Klemm and S. Horn,
%J. Phys. Condense. Matter, {\bf 13}, 8497 (2001).

%\bibitem{Isoda}
%M. Isoda and S. Mori, J. Phys. Soc. Jpn. {\bf 69}, 1509 (2000).

%\bibitem{Fujimoto}
%S. Fujimoto, Phys. Rev. B {\bf 64}, 085102 (2001).

%\bibitem{Tsunetsugu}
%H. Tsunetsugu, J. Phys. Soc. Jpn. {\bf 71}, 1844 (2002).

%\bibitem{Yamashita}
%Y. Yamashita and K. Ueda, 
%Phys. Rev. B {\bf 67}, 195107 (2003).

%\bibitem{Hopkinson}
%J. Hopkinson and P. Coleman,
%Phys. Rev. Lett. {\bf 89}, 267201 (2002).

%%%%%%%%%%%%%%%%

%% LiV2O4 theory

%\bibitem{AnisimovLiVO}
%V. I. Anisimov, {\it et al.}, Phys. Rev. Lett. {\bf 83}, 364 (1999).

%\bibitem{Kusunose}
%H. Kusunose, {\it et al.}, 
%H. Kusunose, S. Yotsuhashi, and K. Miyake,
%Phys. Rev. B {\bf 62}, 4403 (2000).

%\bibitem{Arita}
%R. Arita, K. Held, A. V. Lukoyanov, and V. I. Anisimov, cond-mat/0701509

%% (Ca,Sr)2RuO4
\bibitem{Nakatsuji}
S. Nakatsuji, {\it et al.}, Phys. Rev. Lett. {\bf 90}, 137202 (2003).

%% LaNiO %%
\bibitem{Sreedhar}
K. Sreedhar {\it et al.}, J. Solid State Chem. {\bf 110}, 208 (1994); 
Z. Zhang {\it et al.}, {\it ibid.} {\bf 108}, 402 (1994); {\bf 117}, 236 (1995).

\bibitem{Kobayashi}
Y. Kobayashi, {\it et al.},
%Y. Kobayashi, S. Taniguchi, M. Kasai, M. Sato, T. Nishioka, and
%M. Kontani, 
J. Phys. Soc. Jpn. {\bf 65}, 3978 (1996)











%% (Ca,Sr)2RuO4 OSMT
\bibitem{Anisimov}
V. I. Anisimov {\it et al.}, Eur. Phys. J. B {\bf 25}, 191 (2002).

\bibitem{KogaOSMT}
A. Koga, {\it et al.},
%A. Koga, N. Kawakami, T.M. Rice, and M. Sigrist, 
Phys. Rev. Lett. {\bf 92}, 216402 (2004);
%A. Koga, K. Inaba, and N. Kawakami, 
Prog. Theor. Phys. Suppl. {\bf 160}, 253 (2005).

%% (Ca,Sr)2RuO4
\bibitem{KogaV}
A. Koga, {\it et al.},
%A. Koga, N. Kawakami, T.M. Rice, and M. Sigrist, 
Phys. Rev. B {\bf 72}, 045128 (2005).

\bibitem{Medici}
L. de' Medici, {\it et al.},
%L. de' Medici, A. Georges, and S. Biermann, 
Phys. Rev. B {\bf 72}, 205124 (2005).

\bibitem{previous}
A. Koga, {\it et al.},
%A. Koga, N. Kawakami, R. Peters, and Th. Pruschke,
Phys. Rev. B {\bf 77}, 045120 (2008).
%cond-mat/0708.1765.

\bibitem{Buneman}
J. B\"unemann, {\it et al.},
%J. B\"unemann, D. Rasch, and F. Gebhard,
J. Phys. Condens. Matter, {\bf 19}, 436206 (2007).
%%%%%% OSMT %%%%%%%
%\bibitem{Liebsch}
%A. Liebsch, Europhys. Lett. {\bf 63}, 97 (2003).

%\bibitem{OSMT}
%A. Liebsch,  
%Phys. Rev. Lett. {\bf 91}, 226401 (2003);
%C. Knecht, N. Bl\"umer and P. G. J. van Dongen, 
%Phys. Rev. B {\bf 72}, 081103 (2005);
%A. R\"uegg, M. Indergand, S. Pilgram and M. Sigrist, 
%Eur. Phys. J. B {\bf 48}, 55 (2005);
%M. Ferrero, F. Becca, M. Fabrizio and M. Capone, 
%Phys. Rev. B {\bf 72}, 205126 (2005)
%R. Arita and K. Held, Phys. Rev. B {\bf 72}, 201102(R) (2005).

%\bibitem{Inaba}
%K. Inaba, A. Koga, S. I. Suga, and N. Kawakami,
%Phys. Rev. B {\bf 72}, 085112 (2005);
%J. Phys. Soc. Jpn. {\bf 74}, 2393 (2005);
%K. Inaba and A. Koga,
%Phys. Rev. B {\bf 73}, 155106 (2006).


%%%%%%% PAM
\bibitem{Grewe91}
N. Grewe and F. Steglich, in
  \emph{Handbook on the Physics and Chemistry of Rare
  Earths}, edited by K. A. Gschneidner, Jr. and L. Eyring
  (North-Holland, Amsterdam, 1991), vol. 14, p. 343.

\bibitem{grewe_ssc} 
N. Grewe, Solid State Commun.\ {\bf 50}, 19 (1984);
K. Yamada, and K. Yoshida, in \emph{Theory of Heavy Fermions and Valence
Fluctuations}, edited by T. Kasuya and T. Saso (Springer, Berlin, 1985);
Th. Pruschke, and N. Grewe, Z.\ Phys.\ B {\bf 74}, 439 (1989).

\bibitem{Hewson} A.C. Hewson, \emph{The Kondo Problem to Heavy
  Fermions}, Cambridge University Press (Cambridge, 1993).

%\bibitem{Yamada}
%K. Yamada, {\it et al.},
%K. Yamada, K. Yosida, and K. Hanzawa, 
%Prog. Thero, Phys. Suppl.
%{\bf 108}, 141 (1992).

%%%%%%%%%% Double exchange
\bibitem{Zener}
C. Zener, Phys. Rev. {\bf 82}, 4031 (1951).

\bibitem{Anderson}
P. W. Anderson and H. Hasegawa, Phys. Rev. {\bf 100}, 675 (1955).

\bibitem{Kubo}
K. Kubo and N. Ohata, J. Phys. Soc. Jpn. {\bf 33}, 21 (1975).

\bibitem{Furukawa}
N. Furukawa, J. Phys. Soc. Jpn. {\bf 64}, 2734 (1995).

%%%  DMFT %%%%%%%%%%%%%%%%%%%%%%%
\bibitem{Metzner}
W. Metzner and D. Vollhardt, Phys. Rev. Lett. {\bf 62}, 324 (1989).

\bibitem{Muller}
E. M\"uller-Hartmann, Z. Phys. B {\bf 74}, 507 (1989).

\bibitem{Georges}
A. Georges, {\it et al.},
%A. Georges, G. Kotliar, W. Krauth and M. J. Rozenberg,
Rev. Mod. Phys. {\bf 68}, 13 (1996).

\bibitem{Pruschke}
T. Pruschke, {\it et al.},
%T. Pruschke, M. Jarrell, and J. K. Freericks, 
Adv. Phys. {\bf 42}, 187 (1995).


\bibitem{Chitra}
R. Chitra and G. Kotliar, Phys. Rev. Lett. {\bf 83}, 2386 (1999).

\bibitem{Momoi}
T. Momoi and K. Kubo, Phys. Rev. B {\bf 58}, R567 (1998).

\bibitem{Zitzler}
R. Zitzler, {\it et al.},
%R. Zitzler, Th. Pruschke and R. Bulla, 
Eur. Phys. J. B 27, 473 (2002).

\bibitem{PetersNew}
R. Peters and Th. Pruschke,
Phys. Rev. B {\bf 76}, 245101 (2007).


\bibitem{Schork}
T. Schork and S. Blawid, Phys. Rev. B {\bf 56}, 6559 (1997).

%% PAM Sato
\bibitem{Sato}
R. Sato, {\it et al.},
%R. Sato, T. Ohashi, A. Koga, and N. Kawakami, 
J. Phys. Soc. Jpn. {\bf 73}, 1864 (2004).


%%% NRG 
\bibitem{NRG}
H. R. Krishna-murthy, {\it et al.},
%H. R. Krishna-murthy, J. W. Wilkins, and K. G. Wilson,
Phys. Rev. B {\bf 21}, 1003 (1980).

\bibitem{nrg_rmp} 
R. Bulla, {\it et al.},
%R. Bulla, T. Costi, and Th. Pruschke, 
cond-mat/0701105 (2007).

\bibitem{OSakai}
O. Sakai and Y. Kuramoto, Solid State Comm. {\bf 89}, 307 (1994).

\bibitem{BullaNRG}
R. Bulla, Phys. Rev. Lett. {\bf 83}, 136 (1999).

\bibitem{anders} 
F.B. Anders, and A. Schiller, 
Phys. Rev. B {\bf 74}, 245113 (2006).

\bibitem{Peters}
R. Peters, {\it et al.}, 
%R. Peters, Th. Pruschke, and F. B. Anders, 
Phys. Rev. B {\bf 74}, 245114 (2006).

\bibitem{weichselbaum:06}
A. Weichselbaum and J. von Delft, 
Phys. Rev. Lett. {\bf 99}, 076402 (2007).



%% Field induced phase transition
\bibitem{Ohashi}
T. Ohashi, {\it et al.},
%T. Ohashi, A. Koga, S. I. Suga and N. Kawakami, 
Phys. Rev. B {\bf 70}, 245104 (2004).


\bibitem{Milat}
I.Milat, {\it et al.},
%I.Milat, F. Assaad, and M. Sigrist, 
Eur. Phys. J. B {\bf 38}, 571 (2004).

\end{thebibliography}
\end{document}